\documentclass[twocolumn,english,xlinenumbers,aps,twocolumn]{revtex4-1}
\usepackage[T1]{fontenc}
\usepackage[utf8]{inputenc}
\usepackage{color}
\usepackage{babel}
\usepackage{verbatim}
\usepackage{amsmath}
\usepackage{graphicx}
\usepackage[unicode=true,pdfusetitle,
 bookmarks=true,bookmarksnumbered=false,bookmarksopen=false,
 breaklinks=false,pdfborder={0 0 0},pdfborderstyle={},backref=false,colorlinks=true]
 {hyperref}
\hypersetup{
 citecolor=blue}
\usepackage{breakurl}
\usepackage{textcomp}
\usepackage{lmodern}



\begin{document}

\title{Temporal Localized Structures in mode-locked Vertical External-Cavity Surface-Emitting Lasers}

\author{P. Camelin$^{1}$, C. Schelte$^{2,3}$, A. Verschelde$^{1}$, A. Garnache$^{4}$, G. Beaudoin$^{5}$, I. Sagnes$^{5}$, G. Huyet$^{1}$, J. Javaloyes$^{2,*}$, S. V. Gurevich$^{3,6}$, M. Giudici$^{1}$ }

\affiliation{$^{1}$ Université Côte d'Azur, Centre National de la Recherche Scientifique, Institut de Physique de Nice, 06560 Valbonne, France}
\affiliation{$^{2}$ Departament de Física, Universitat de les Illes Balears, C/ Valldemossa km 7.5, 07122 Palma de Mallorca, Spain}
\affiliation{$^{3}$ Institute for Theoretical Physics, University of Münster, Wilhelm-Klemm-Str. 9, 48149 Münster, Germany}
\affiliation{$^{4}$ Institut d'Electronique et des Systémes, UMR5214, Centre National de la Recherche Scientifique, University of Montpellier, France}
\affiliation{$^{5}$ Centre de Nanosciences et de Nanotechnologies, Centre National de la Recherche Scientifique, Université Paris Sud, 91460 Marcoussis}
\affiliation{$^{6}$ Center for Nonlinear Science (CeNoS), University of Münster, Corrensstraße 2, 48149 Münster, Germany}
\affiliation{$^{*}$ Corresponding author: 	Julien.Javaloyes@uib.es}

\begin{abstract}
Temporal Localized States (TLSs) are individually addressable structures traveling in optical resonators.
They can be used as bits of information and to generate frequency combs with tunable spectral density.
We show that a pair of specially designed nonlinear mirrors, a \textonehalf{} Vertical-Cavity Surface-Emitting
Laser and a Semiconductor Saturable Absorber, coupled in self-imaging conditions, can lead to the generation
of such TLSs. Our results indicate how a conventional passive mode-locking scheme can be adapted to provide
a robust and simple system emitting TLSs and it paves the way towards the
observation of three dimensions confined states, the so-called light bullets.
\end{abstract}

\maketitle

\section{introduction}

The possibility of using optical Localized States (LSs) in resonators to code and process information bits has attracted a great deal of attention in the last twenty years \citep{1172836,AFO-AAM-09,1172836,BCE-APX-17}.
LSs are stable solutions of dissipative systems characterized by a correlation range much shorter than the size of the underlying system. They can be individually switched on and off by a local
perturbation. Two conditions appear to be necessary for a system to exhibit LSs: i) to have a large aspect-ratio, i.e. to be extended enough that boundary conditions do not play a relevant role and ii) to feature a
bistable response, typically between a pattern and a homogeneous state. Localized Structures were first observed in the transverse section of a broad-area Vertical Cavity Surface Emitting Laser (VCSEL) where they appear
as localized beams of light (spatial LSs) \citep{BTB-NAT-02,TAF-PRL-08,GBG-PRL-08}. More recently, \emph{temporal} LSs (TLSs) have been reported along the longitudinal dimension of an optical resonator where they appear
as individually addressable circulating pulses \citep{LCK-NAP-10,HBJ-NAP-14}. In semiconductor lasers, TLSs have been predicted theoretically in the context of passive mode-locking (PML) \citep{MJB-PRL-14}. It was
shown that in the limit where the cavity round-trip $\tau$ is longer than the gain recovery $\tau_{g}$ and for large modulation of the losses induced by the absorber, the bifurcation towards the fundamental PML becomes
subcritical, thus leading to bistability between the pulsating and the off solutions. This prediction was experimentally implemented by using an electrically-pumped broad-area VCSEL coupled to a resonant semiconductor
saturable absorber mirror (SESAM) \citep{MJB-PRL-14,MJC-JSTQE-16,CJM-PRA-16}. The use of a VCSEL as an amplifier made it problematic to match a necessary condition for obtaining a stable PML regime, namely that the gain
section saturation fluence should be larger than the absorber's one. This difficulty was circumvented by imaging the VCSEL far-field transverse profile onto the SESAM in a self-collimating scheme \citep{MJB-JSTQE-15}.
This peculiar configuration favored a tilted wave emission from the \emph{whole} surface of the VCSEL, the latter being imaged as a \emph{single point} in the SESAM. Although this scheme allowed saturating the SESAM
very efficiently, it also generated complex transverse dynamics coupling the whole transverse profile of the broad-area VCSEL. A simpler and more robust design for generating TLSs would rather be based on a
self-imaging condition between the gain and the saturable absorber mirror.

In this paper we describe the implementation of temporal LSs in a self-imaging passive mode-locked system based on
an optically-pumped Vertical External Cavity Surface Emitting Laser (VECSEL). Both the gain mirror (also called \textonehalf{} VCSEL)
and the SESAM have been specially designed to match the parameter requirements for achieving localization in this widely used configuration
for generating PML at GHz rate \citep{garnache02,tropper04,KT-PR-06}. We employ a first principle model \citep{MB-JQE-05,MJB-JSTQE-15}
to identify the key design parameters and, in particular, to achieve robust multistability together with an accessible
lasing threshold. The localization regime is disclosed by the coexistence in the parameter space of a wide range of pulsating solutions
with a number of pulses per round-trip spanning from zero (trivial solution) up to a maximal number $N$, with $N$ being the order of the
harmonic PML solution appearing at the lasing threshold \citep{MJB-PRL-14}.

Our results pave the way towards the observation of spatio-temporal localized structures, also called light bullets (LBs) \citep{WD-OPN-02}. These objects have been actively sought in conservative systems in the last
three decades, yet only fading LBs were observed experimentally so far \citep{MEK-PRL-10}. Recently, \emph{dissipative} LBs were predicted in PML resonators featuring a large aspect ratio both in the temporal domain
and in the spatial domain \citep{J-PRL-16,GJ-PRA-17}. Since the former condition implies an extended longitudinal dimension of the resonator, only a self-imaging configuration between the gain and the absorbing mirrors
can impede diffraction to affect detrimentally the Fresnel number of the cavity \citep{GBG-PRL-08}. In this regime, the large spatial aspect ratio can be achieved by pumping a large area of the gain mirror. In this
work, even if the optically pumped area is too small for achieving a large Fresnel number, the existence of temporal LSs in a self-imaging cavity is a decisive step towards the experimental observation of LBs.
Moreover, the temporal localization regime we present offers important advantages as compared to conventional PML since it enables arbitrary low repetition rates and the generation of arbitrary pulse patterns. In terms
of frequency comb generation, the coexistence of a large set of regular pulsating solutions with a number of pulses $N$ comprised in the interval $N\in\left[1:\tau/\tau_{g}\right]$ allows to change the frequency
density of the comb, which is very interesting for adapting the comb characteristics to the spectrum under analysis. These characteristics may be attractive for several applications, e.g., time-resolved spectroscopy,
pump-probe sensing, optical code division multiple access communication networks and LIDAR.

%\section*{2. VECSEL design}

Passive mode-locking is usually implemented using a VECSEL by closing the external cavity with
a concave mirror which selects a single transverse mode and limits losses \citep{garnache02,tropper04,KT-PR-06}.
Our requirements on large spatial and temporal aspect-ratios leads us to consider a linear resonator
delimited by the gain mirror  facing the SESAM in a self-imaging configuration. This plane-parallel
extended resonator implies additional losses with respect to conventional PML schemes. In addition, light must be extracted
by using an intra-cavity AR coated glass window which induces additional losses (2\%\,reflection).
Accordingly, the gain mirror needs to be designed for providing a gain level capable of overcoming
such an unusually high level of losses. For the same reasons the SESAM needs to be designed to have the
smallest amount of unsaturable losses. We matched these requirements by realizing a gain
mirror based on a GaAs substrate with 6 strain-balanced InGaAs/GaAsP quantum wells (QWs)
designed for optical pumping and emitting at $1.06\,\mu$m \citep{LMB-OE-10,CZF-AO-18}.
The QW gain material is operated under large carrier excitation to reduce carrier lifetime
below $\tau_{g}=3\,$ns. The high reflectivity bottom Bragg mirror is made of 31~AlAs/AlGaAs layers
exhibiting good thermal conductivity for  reduced thermal impedance. The high level of external cavity
losses was compensated by strengthening the residual micro-cavity effect by adding 3 Bragg's pair
on the top of the gain device, which increases the confinement factor of the field radiation
to a value of $\Gamma=12$, thus leading to a cavity bandwidth of $\Delta\lambda_{G}=3.5\,$nm (FWHM),
a gain saturation fluence $F_{G}=10\,\mu$J/cm$^{2}$ and a gain level at maximum pumping
of $130\,\%$. While a higher confinement factor is beneficial to withstand a
relatively high level of losses, it is necessary to compare $F_{G}$ with the SESAM
saturation fluence ($F_{B}$) in order to fulfill the requirement $s=F_{G}/F_{B}>1$  which ensures stable PML.
We designed a SESAM featuring a single strained InGaAs/GaAs QW located at $\sim1\text{--}2\,$ nm
from the external surface \citep{garnache02}. The fabrication process was optimized
for minimizing non saturable losses (less than $0.3\,\%$). The SESAM's carriers recombination
time is of the order of $50\,$ps. The back mirror is highly reflective ($R_{b}\simeq99.9\,\%$) while
the top mirror reflectivity has been varied producing different samples in order to match the parameter
requirements according to the indications of the model \cite{GJ-PRA-17}.
In the sample used for this paper the micro-cavity was defined by the air/semiconductor
interface and its finesse was increased by adding a top dielectric Bragg mirror formed
by a pair of Si/SiN layers, leading to a wavelength bandwidth of $\Delta\lambda_{B}=15\,$ns (FWHM),
$F_{B}=2\,\mu$J/cm$^{2}$ and a modulation of the reflectivity at resonance of $A=25\,\%$.

The gain mirror is optically pumped at $780\,$nm by a single transverse mode diode laser.
The pump beam is incident onto the gain mirror at the Brewster angle ($\theta=74^{o}$) and it
has a Gaussian shape with $50\,\mu$m diameter (FWHM). Both the gain mirror and SESAM devices
are thermally stabilized and controlled by a Peltier cell.
The gain mirror output is collimated by using an aspherical collimator having a focal length
of $8\,$mm and a numerical aperture of 0.5. The same collimator is placed in front of the SESAM.
We inserted into the cavity two biconvexe lenses of $5\,$cm focal length in order to
obtain a self-imaging configuration between the gain mirror and the SESAM emitting surfaces.
The cavity length was set to $L=160\,$cm, which implies a cavity round-trip time of $\tau=10.7\,$ns.

%\section*{3. Experimental Results}

\begin{figure}
\includegraphics[width=1.2\columnwidth]{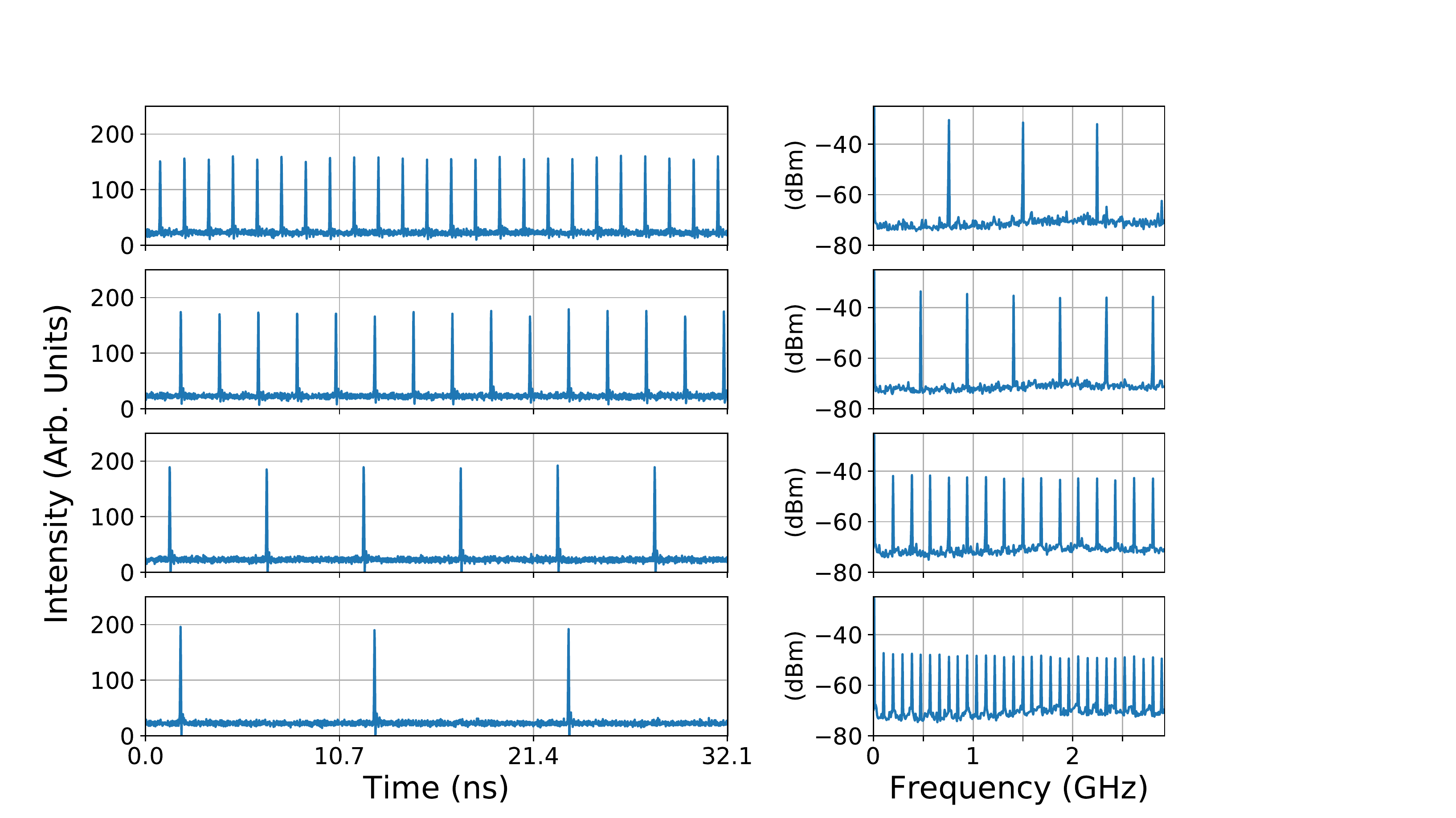}
\caption{Time series (left) and Power spectra (right) of some of the coexisting emission states
at $P=206\,$mW. From the top to the bottom: 8,\,5,\,2, and 1 pulse(s) per round-trip. \label{Time-series-160}}
\end{figure}

Thermal control of the gain mirror and SESAM substrates allows to tune
the resonance condition between the two residual micro-cavities, thus
tuning the amount of saturable losses experienced by the pulse circulating
in the cavity. We define the detuning between the two cavities
as $\delta\lambda=\lambda_{A}-\lambda_{G}$. At $T=30^{\circ}$ the
resonances are located at $\lambda_{A}=1066.9\,$nm for the SESAM
and $\lambda_{G}=1064.5\,$nm for the gain mirror pumped at $P=300\,$mW.
It turns out that the laser threshold was beyond the maximal pumping
power available ($P_{M}<325\,$mW) if the two resonances were brought
too close from each other since in this case, the system is forced
to operate with the maximal level of unsaturated losses. The lasing
threshold could be reached only if $\delta\lambda>1.5\,$nm, at this
limit value the saturable losses are decreased roughly to $A\sim19\%$.
With beam sampler transmission of $T_{b}=98\,\%$ this corresponds
to a unsaturated reflectivity of the gain mirror at threshold of $R_{G}=\left(R_{B}T_{b}^{2}\right)^{-1}=129\,\%$.

The limit setting at $\delta\lambda=1.5\,$nm is also the one ensuring
the bistable response of the system on the largest pumping range.
While the threshold power decreases as $\delta\lambda$ is increased,
the bistability range becomes also narrower. The detuning was finally
set at $\delta\lambda=3.4\,$nm for which $A\sim17\%$ and $R_{G}=125\,\%$.
a value ensuring a sufficiently large bistable response and a threshold
pump value that remains well below to the maximal pump power limit.

\begin{figure}
\includegraphics[viewport=20bp 0bp 620bp 420bp,clip,height=4.5cm]{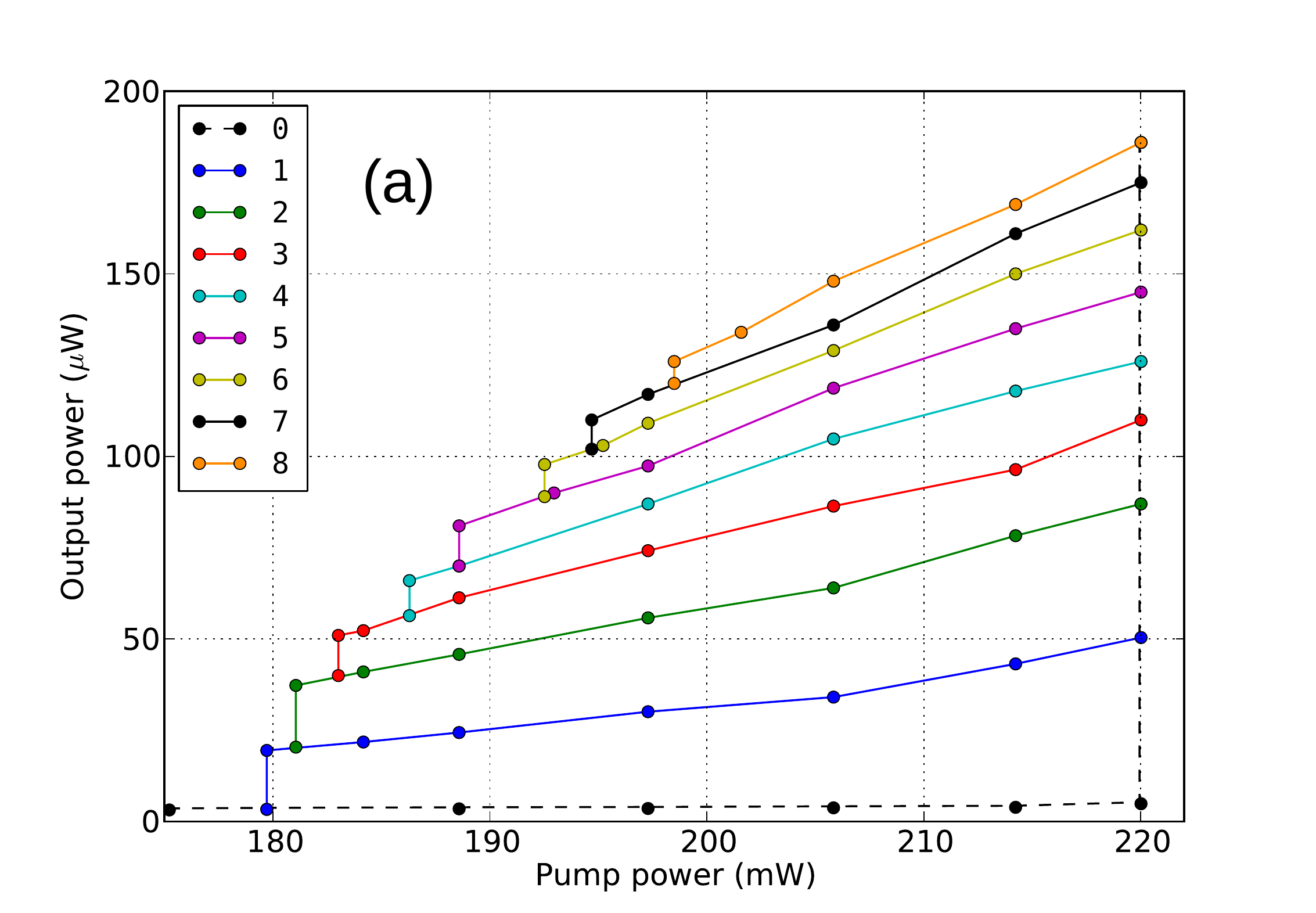}%
\includegraphics[viewport=0bp 30bp 270bp 540bp,clip,height=4.5cm]{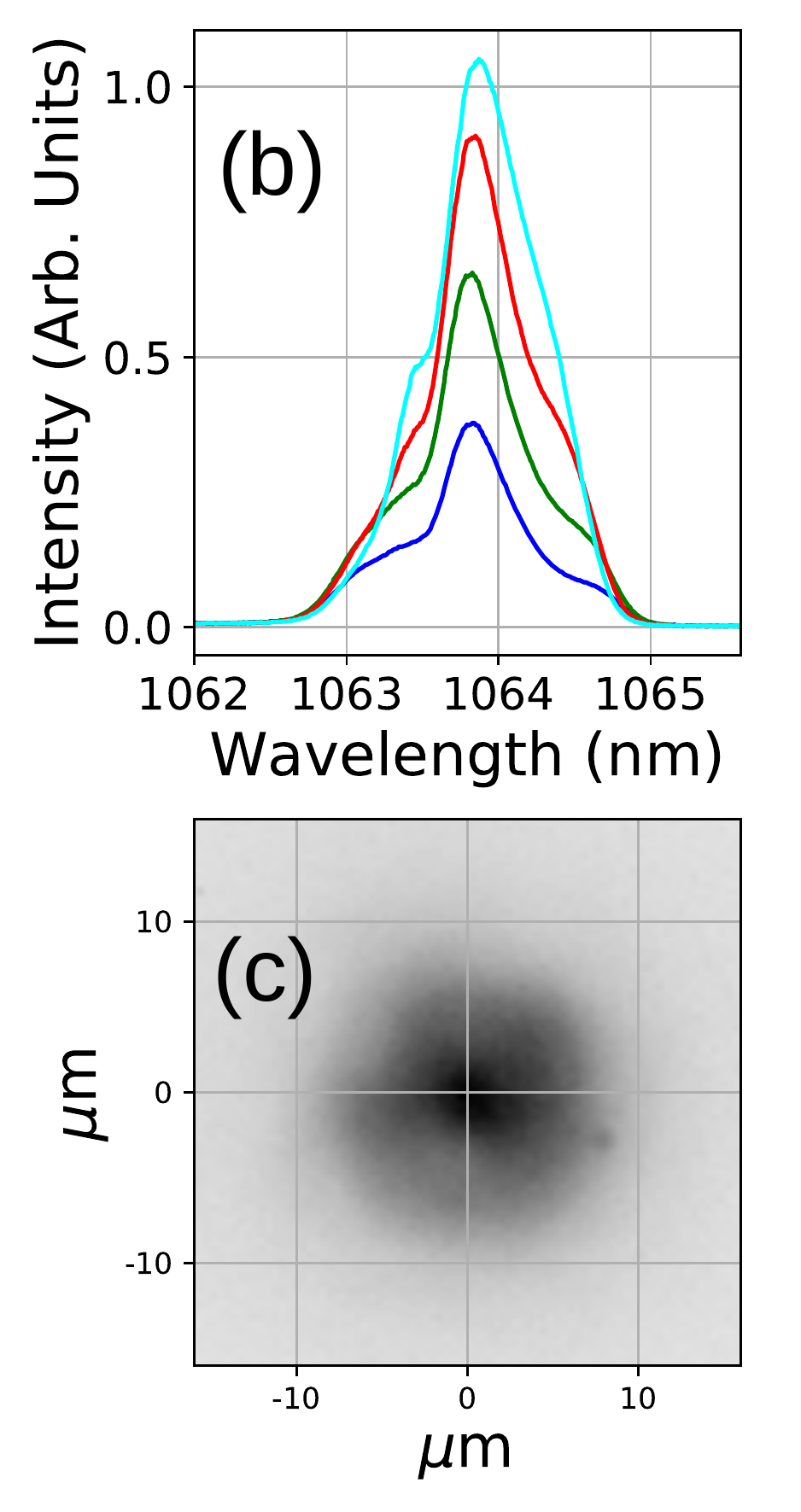}
\caption{(a) Bifurcation diagram of the pulsed emission present into the system below the laser threshold ($P_{th}=220\,$mW) from the solution with $N=0$ to the solution with $N=8$ pulses per round-trip. (b) Optical
spectra corresponding to the emission of $N=1$ to $N=4$ pulses $P=206\,$mW. (c) Near-field image of the VECSEL emission.} \label{Bif-diag-160}
\end{figure}

An example of several coexisting pulsating solutions are shown in Fig.~\ref{Time-series-160}.
A regime with $N=8$ pulses per round-trip was observed when the pumping
intensity was brought above $P_{th}=220\,$mW. Because at this pumping value
the off solution becomes unstable, this corresponds to the VECSEL threshold value.
The stability of the $N=8$ solution appearing at threshold can be tested
by decreasing the pumping power. It is observed that this regime is stable
down to $P_{s}^{(8)}=198\,$mW where it gives way to the pulsed solution
with  $N=7$, see Fig.~\ref{Bif-diag-160}~(a). When the jump occurs the
stability of the $N=7$ solution can be tested by increasing the pump
power until $P_{th}$ and then decreasing it again. When increasing the
pump power, one needs to take care not to cross the threshold value
otherwise the system will jump back to the $N=8$ solution. We verified
that the $N=7$ solution is stable for $P\in\left[P_{sn}^{\left(7\right)}:P_{th}\right]$
with $P_{sn}^{\left(7\right)}=194\,$mW. When the pumping power is
decreased below $P_{sn}^{\left(7\right)}$ we observe that a new
solution with $N=6$ appears whose stability is similarly assessed
for $P\in\left[P_{sn}^{\left(6\right)}:P_{th}\right]$ with
$P_{sn}^{\left(6\right)}=192\,$mW. The same procedure allows for
testing the full set of solutions branches and leads to the full
bifurcation diagram reported in Fig.~\ref{Bif-diag-160}~a)
featuring a staircase structure with a generalized multistability
between all the pulsed emission states. This demonstrates that
the system is operated in the localization regime and that up to $N=8$
pulses can be written in the cavity and erased individually.
The RF combs in the power spectra are composed by lines whose width
is less than $1\,$kHz (FWHM) which is the RF bandwidth of our
instrument and $2\,$kHz FWHM at $-10\,$dB. The intensity profile of
the TLSs could not be resolved by our detection system whose
bandwidth is $34\,$GHz. The optical spectrum analyzer trace
indicates a FWHM span of $\delta\lambda=0.7\,$nm that is independent
of $N$, as shown in Fig.~\ref{Bif-diag-160}~(b). The near-field emission
of the system is depicted in Fig.~\ref{Bif-diag-160}~(c) showing the
presence of single spot whose width is $w_{p}=15\,\mu$m (FWHM)
and at the center of the pumped region.

%\section*{Theoretical Results}\label{sec:Theoretical-Results}

Our theoretical approach follows the method developed in \citep{MB-JQE-05,MJB-JSTQE-15}
in which the QW active region is considered as infinitely thin and
located at an antinode of the field, and the wave equation is solved
exactly, in Fourier domain, within the linear sections of the micro-cavities.
The dynamical model for the intra-cavity fields $E_{j}$, where the
indexes $j=1$ and $j=2$ denote the VECSEL and the SESAM, respectively, reads

\begin{eqnarray}
\kappa_{1}^{-1}\dot{E}_{1} & = & \left[\left(1-i\alpha_{1}\right)N_{1}-1\right]E_{1}+h_{1}Y_{1},\label{eq:F1}\\
\kappa_{2}^{-1}\dot{E}_{2} & = & \left[\left(1-i\alpha_{2}\right)N_{2}-1+i\delta\right]E_{2}+h_{2}Y_{2},\label{eq:F2}\\
\dot{N}_{1} & = & \gamma_{1}\left(J_{1}-N_{1}\right)-\left|E_{1}\right|^{2}N_{1},\label{eq:F3}\\
\dot{N}_{2} & = & \gamma_{2}\left(J_{2}-N_{2}\right)-s\left|E_{2}\right|^{2}N_{2}.\label{eq:F4}
\end{eqnarray}

Here, the photon lifetimes are $\kappa_{j}^{-1}$, the detuning between the
two cavities is $\delta$, $\alpha_{j}$ denote the linewidth enhancement
factors and $N_{j}$ are the two population inversions whose lifetimes are
$\gamma_{j}^{-1}$, $j=1,\,2$. The fields injected into the micro-cavities are $Y_{j}$
with a coupling efficiency given by the parameters $h_{j}$.
The two cavities mutually inject each other and their outputs consist in
a superposition between the reflected and emitted fields.
Considering the time of flight in the cavity $\tau$ as well as the presence of the beam
sampler with amplitude transmission $t_{bs}$, the link between the two cavities,
taking into account all the multiple reflections,
is given by two Delay Algebraic Equations that read
\begin{eqnarray}
Y_{1,2}\left(t\right) & = & \sqrt{T_{b}}\left[E_{2,1}\left(t-\tau\right)-Y_{2,1}\left(t-\tau\right)\right].\label{eq:F5}
\end{eqnarray}
% \begin{eqnarray}
% Y_{j}\left(t\right) & = & \sqrt{T_{b}}\left[E_{3-j}\left(t-\tau\right)-Y_{3-j}\left(t-\tau\right)\right].\label{eq:F5}
% \end{eqnarray}
We stress that the cavity enhancement factors due to the highly reflective
mirrors has been scaled out using Stokes relations, making  $E_{j}$
and $Y_{j}$ of the same order of magnitude. This scaling has
the additional advantage of simplifying the input-output relations
given by Eq.~\ref{eq:F5}. The minus sign represents the $\pi$ phase
shift of the incoming field $Y$ upon reflection from the top Bragg
mirror. Finally, we stress that the saturation parameter $s$ in
Eq.~\ref{eq:F4} contains both the ratio of the active material
differential gains and of the cavity enhancement factors.

We set the photon lifetimes as $\left(\kappa_{1}^{-1},\kappa_{2}^{-1}\right)=\left(343,\,80\right)\,$fs
which corresponds to $3.5\,$nm and $15\,$nm for the FWHM of the resonance, respectively.
The gain and absorber linewidth enhancement factors and carrier lifetimes
are $\left(\alpha_{1},\alpha_{2}\right)=\left(2.5,1\right)$ and
$\left(\gamma_{1}^{-1},\gamma_{2}^{-1}\right)=\left(800,\,50\right)\,$ps.
The bias in the gain is varied in the interval $J_{1}\in\left[0\,0.1\right]$
and we set $h_{1}=2$. We remark that $J_{1}=0.065$ corresponds to
the maximal allowed bias experimentally as it corresponds to a peak
reflectivity of $129\,\%$. The bias in the absorber is set to $J_{2}=-0.07$
and we set $h_{2}=1.9985$ leading to unsaturated reflectivity of $74\,\%$
and non saturable losses of $0.3\,\%$. The saturation parameter is $s=5$ and $T_b=t_{bs}^2=0.99$.

\begin{figure}
\centering{}\includegraphics[width=1\columnwidth]{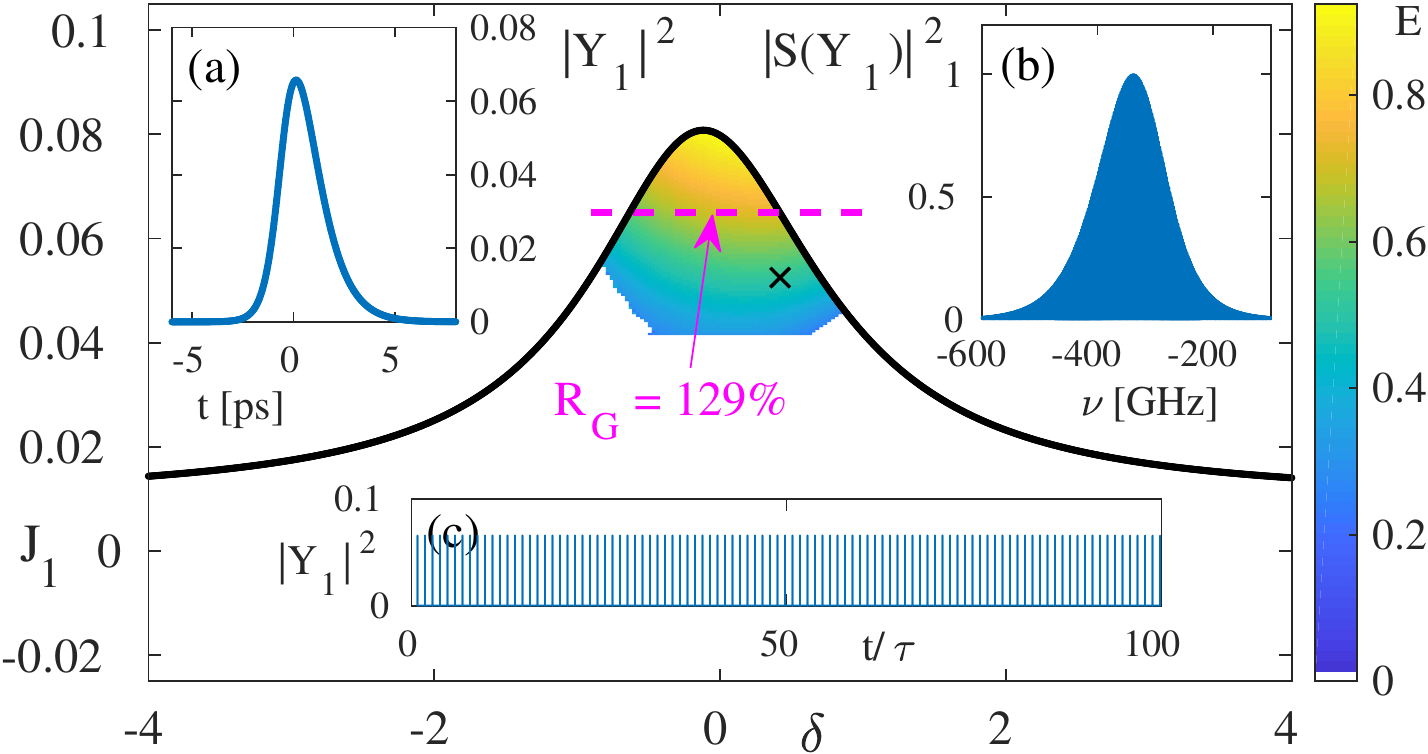}
\caption{(a) Two-dimensional bifurcation diagram showing the region of stable existence
of the TLSs given by Eqs.~(\ref{eq:F1}-\ref{eq:F5}) as a function of the forward bias $J_{1}$
and the detuning $\delta$. The color code represent the pulse energy, the black and dashed pink lines
represent the lasing threshold and an unsaturated gain of $129\%$, respectively.
(a,b) Pulse shape and Optical spectrum obtained for $\left(J_{1},\delta\right)=\left(0.0525,0.42\right)$ and (c) Temporal trace.}
\label{fig:TLS_diag}
\end{figure}

As we operate in the long delay limit, the recently developed functional mapping for PML is the most adapted method \citep{SJG-OL-18}. This approach is based upon computing only the so-called \textit{fast stage} in the
vicinity of the pulse where stimulated emission is dominant, while using the analytical solution of the dynamics during the \textit{slow stages} in-between pulses, see  \citep{SJG-OL-18} for details. This allowed us to
perform multidimensional parameter maps and we present some of our results in Fig.~\ref{fig:TLS_diag}, where we scanned the region of bistability by direct numerical continuation as a function of the bias current
$J_{1}$ and the detuning between the two cavities $\delta$. We see that for large detuning, the lasing threshold is easily reached but the absorber is off resonance and no PML occurs as there is no modulation of the
SESAM. Oppositely, the bistability region is maximal for low values of $\delta$ for which the modulation of the SESAM is maximal. For $\delta\sim 0$, the threshold is out of reach and TLSs could only be triggered by
external perturbations, i.e. by injecting optical pulses. This justifies the need to operate with some minimal amount of detuning, which is in excellent agreement with the experiment. We notice in
Fig.~\ref{fig:TLS_diag}~(a) that a value of  $\delta=0.42$ (i.e., $\delta \lambda=3.15\,$nm) represents a good compromise as the threshold can be reached while conserving a wide bistable region. We noticed that the extend and
the shape of the bistable region depends strongly on $\left(\alpha_{1},\alpha_{2}\right)$. They will be the topic of further studies. Finally, we present in Fig.~\ref{fig:TLS_diag}~(a, b) the detail of the pulse
characteristics calculated at $(J_1,\,\delta)=(0.0525,\,0.42)$, whereas in Fig.~\ref{fig:TLS_diag}~(c) the temporal trace is depicted. We find a temporal and spectral width (FWHM) of $2.33\,$ps and $140\,$ GHz (i.e., $0.17\,$nm), respectively, leading to a time bandwidth product of 0.32. We explain such a low value by the asymmetry of pulse temporal profile, see Fig.~\ref{fig:TLS_diag}~(a). Finally, the fact that simulated pulses are spectrally more narrow than experimentally can be ascribed to the fact that we did not consider the transverse spatial dynamics, which can lead to additional dispersion.

%
%c=3e8; l0=1064e-9; 0.15e-9*pi*c/l0^2
%

In conclusion, we presented how TLSs can be obtained in self-imaging VECSELs. The experimental realization was achieved by engineering micro-cavities with parameters that match the uncommon requirements for TLSs
according to the indications of a first principle model based on delay algebraic equations. We identified how the detuning between the two micro-cavities allows to combine robust multi-stability and low threshold.
We foresee that further optimization such as a SESAM with lower bandwidth and a lower saturation energy would yield an even greater bistable region. Our results have been obtained by pumping a portion of the gain mirror
which is too small for achieving spatial localization. The evidence of temporal localization is an encouraging indication that Light Bullets may be experimentally observed in self-imaging VECSELs by increasing the pumped region area. To this aim, several technological issues are currently addressed, namely in terms of thermal management and pump beam shaping.

\section*{Funding information}
MINECO Project COMBINA (TEC2015-65212-C3-3-P).

Platforme OPTIMAL funded by CNRS and FEDER PACA. P. Camelin acknowledge CNRS and R\'egion PACA (Emplois Jeunes Doctorants).


\begin{thebibliography}{10}

\bibitem{1172836}
L.A. Lugiato.
\newblock Introduction to the feature section on cavity solitons: An overview.
\newblock {\em Quantum Electronics, IEEE Journal of}, 39(2):193--196, 2003.

\bibitem{AFO-AAM-09}
T.~Ackemann, W.~J. Firth, and G.L. Oppo.
\newblock Chapter 6 fundamentals and applications of spatial dissipative
  solitons in photonic devices.
\newblock In P.~R.~Berman E.~Arimondo and C.~C. Lin, editors, {\em Advances in
  Atomic Molecular and Optical Physics}, volume~57 of {\em Advances In Atomic,
  Molecular, and Optical Physics}, pages 323 -- 421. Academic Press, 2009.

\bibitem{BCE-APX-17}
S.~Barland, S.~Coen, M.~Erkintalo, M.~Giudici, J.~Javaloyes, and S.~Murdoch.
\newblock Temporal localized structures in optical resonators.
\newblock {\em Advances in Physics: X}, 2(3):496--517, 2017.

\bibitem{BTB-NAT-02}
S.~Barland, J.~R. Tredicce, M.~Brambilla, L.~A. Lugiato, S.~Balle, M.~Giudici,
  T.~Maggipinto, L.~Spinelli, G.~Tissoni, T.~Kn{\"o}dl, M.~Miller, and
  R.~J{\"a}ger.
\newblock Cavity solitons as pixels in semiconductor microcavities.
\newblock {\em Nature}, 419(6908):699--702, Oct 2002.

\bibitem{TAF-PRL-08}
Y.~Tanguy, T.~Ackemann, W.~J. Firth, and R.~J\"ager.
\newblock Realization of a semiconductor-based cavity soliton laser.
\newblock {\em Phys. Rev. Lett.}, 100:013907, Jan 2008.

\bibitem{GBG-PRL-08}
P.~Genevet, S.~Barland, M.~Giudici, and J.~R. Tredicce.
\newblock Cavity soliton laser based on mutually coupled semiconductor
  microresonators.
\newblock {\em Phys. Rev. Lett.}, 101:123905, Sep 2008.

\bibitem{LCK-NAP-10}
F.~Leo, S.~Coen, P.~Kockaert, S.P. Gorza, P.~Emplit, and M.~Haelterman.
\newblock Temporal cavity solitons in one-dimensional kerr media as bits in an
  all-optical buffer.
\newblock {\em Nat Photon}, 4(7):471--476, Jul 2010.

\bibitem{HBJ-NAP-14}
T.~Herr, V.~Brasch, J.~D. Jost, C.~Y. Wang, N.~M. Kondratiev, M.~L. Gorodetsky,
  and T.~J. Kippenberg.
\newblock Temporal solitons in optical microresonators.
\newblock {\em Nature Photonics}, 8(2):145--152, 2014.

\bibitem{MJB-PRL-14}
M.~Marconi, J.~Javaloyes, S.~Balle, and M.~Giudici.
\newblock How lasing localized structures evolve out of passive mode locking.
\newblock {\em Phys. Rev. Lett.}, 112:223901, Jun 2014.

\bibitem{MJC-JSTQE-16}
M.~Marconi, J.~Javaloyes, P.~Camelin, D.~Chaparro, S.~Balle, and M.~Giudici.
\newblock Control and generation of localized pulses in passively mode-locked
  semiconductor lasers.
\newblock {\em Selected Topics in Quantum Electronics, IEEE Journal of},
  PP(99):1--1, 2015.

\bibitem{CJM-PRA-16}
P.~Camelin, J.~Javaloyes, M.~Marconi, and M.~Giudici.
\newblock Electrical addressing and temporal tweezing of localized pulses in
  passively-mode-locked semiconductor lasers.
\newblock {\em Phys. Rev. A}, 94:063854, Dec 2016.

\bibitem{MJB-JSTQE-15}
M.~Marconi, J.~Javaloyes, S.~Balle, and M.~Giudici.
\newblock Passive mode-locking and tilted waves in broad-area vertical-cavity
  surface-emitting lasers.
\newblock {\em Selected Topics in Quantum Electronics, IEEE Journal of},
  21(1):85--93, Jan 2015.

\bibitem{garnache02}
A.~Garnache, S.~Hoogland, A.~C. Tropper, I.~Sagnes, G.~Saint-Girons, and J.~S.
  Roberts.
\newblock Sub-500-fs soliton-like pulse in a passively mode-locked broadband
  surface-emitting laser with 100 {mW} average power.
\newblock {\em Applied Physics Letters}, 80:3892--3894, 2002.

\bibitem{tropper04}
A.~C. Tropper, H.~D. Foreman, A.~Garnache, K.~G. Wilcox, and S.~H. Hoogland.
\newblock Vertical-external-cavity semiconductor lasers.
\newblock {\em J. Phys. D: Appl. Phys.}, 37:R75--R85, 2004.

\bibitem{KT-PR-06}
U.~Keller and A.~C. Tropper.
\newblock Passively modelocked surface-emitting semiconductor lasers.
\newblock {\em Physics Reports}, 429(2):67 -- 120, 2006.

\bibitem{MB-JQE-05}
J.~Mulet and S.~Balle.
\newblock Mode locking dynamics in electrically-driven vertical-external-cavity
  surface-emitting lasers.
\newblock {\em Quantum Electronics, IEEE Journal of}, 41(9):1148--1156, 2005.

\bibitem{WD-OPN-02}
Frank Wise and Paolo~Di Trapani.
\newblock Spatiotemporal solitons.
\newblock {\em Opt. Photon. News}, 13(2):28--32, Feb 2002.

\bibitem{MEK-PRL-10}
S.~Minardi, F.~Eilenberger, Y.~V. Kartashov, A.~Szameit, U.~R\"opke,
  J.~Kobelke, K.~Schuster, H.~Bartelt, S.~Nolte, L.~Torner, F.~Lederer,
  A.~T\"unnermann, and T.~Pertsch.
\newblock Three-dimensional light bullets in arrays of waveguides.
\newblock {\em Phys. Rev. Lett.}, 105:263901, Dec 2010.

\bibitem{J-PRL-16}
J.~Javaloyes.
\newblock Cavity light bullets in passively mode-locked semiconductor lasers.
\newblock {\em Phys. Rev. Lett.}, 116:043901, Jan 2016.

\bibitem{GJ-PRA-17}
S.~V. Gurevich and J.~Javaloyes.
\newblock Spatial instabilities of light bullets in passively-mode-locked
  lasers.
\newblock {\em Phys. Rev. A}, 96:023821, Aug 2017.

\bibitem{LMB-OE-10}
A.~Laurain, M.~Myara, G.~Beaudoin, I.~Sagnes, and A.~Garnache.
\newblock Multiwatt---power highly---coherent compact single---frequency
  tunable
  vertical---external---cavity---surface---emitting---semiconductor---laser.
\newblock {\em Opt. Express}, 18(14):14627--14636, Jul 2010.

\bibitem{CZF-AO-18}
Baptiste Chomet, Jian Zhao, Laurence Ferrieres, Mikhael Myara, Germain Guiraud,
  Gr\'{e}goire Beaudoin, Vincent Lecocq, Isabelle Sagnes, Nicholas Traynor,
  Giorgio Santarelli, Stephane Denet, and Arnaud Garnache.
\newblock High-power tunable low-noise coherent source at
  1.06\textmu m based on a surface-emitting semiconductor
  laser.
\newblock {\em Appl. Opt.}, 57(18):5224--5229, Jun 2018.

\bibitem{SJG-OL-18}
C.~Schelte, J.~Javaloyes, and S.~V. Gurevich.
\newblock Functional mapping for passively mode-locked semiconductor lasers.
\newblock {\em Opt. Lett.}, 43(11):2535--2538, Jun 2018.

\end{thebibliography}
\end{document}